\documentclass[aps,10pt,prb,twocolumn,letterpaper,superscriptaddress]{revtex4-2}

\usepackage{graphicx} % Don't include figure files
\usepackage{bm} % bold math
\usepackage{amsmath}

\usepackage{xcolor}

\definecolor{PRLblue}{rgb}{0.18,0.18,0.57}
\usepackage[colorlinks=true,allcolors=PRLblue]{hyperref} % add hypertext capabilities
\graphicspath{ {./}{figs/}}

\usepackage{fancyhdr}
\begin{document}
\title{Mesoscale variations of chemical and electronic landscape on the surface of Weyl semimetal \texorpdfstring{Co$_3$Sn$_2$S$_2$}{Co3Sn2S2} visualized by ARPES and XPS}

\author{Sudheer Anand Sreedhar}
\email{ssreedhar@ucdavis.edu}
\affiliation{Department of Physics and Astronomy, University of California, Davis, CA 95616, USA}

\author{Matthew Staab}
\affiliation{Department of Physics and Astronomy, University of California, Davis, CA 95616, USA}

\author{Mingkun Chen}
\affiliation{Department of Physics and Astronomy, University of California, Davis, CA 95616, USA}

\author{Robert Prater}
\affiliation{Department of Physics and Astronomy, University of California, Davis, CA 95616, USA}

\author{Zihao Shen}
\email{zhshen17@gmail.com}
\affiliation{Department of Physics and Astronomy, University of California, Davis, CA 95616, USA}

\author{Giuseppina Conti}
\affiliation{Advanced Light Source, Lawrence Berkeley National Lab, Berkeley, 94720, USA}

\author{Ittai Sidilkover}
\affiliation{School of Physics and Astronomy, Faculty of Exact Sciences, Tel Aviv University, Tel Aviv-6997801, Israel}
\affiliation{Center for Light-Matter Interaction, Tel Aviv University, Tel-Aviv 6997801, Israel}

\author{Zhenghong Wu}
\email{zhenghow@andrew.cmu.edu}
\affiliation{Department of Physics, Carnegie Mellon University, Pittsburgh, PA, 15213, USA}

\author{Eli Rotenberg}
\affiliation{Advanced Light Source, Lawrence Berkeley National Lab, Berkeley, 94720, USA}

\author{Aaron Bostwick}
\affiliation{Advanced Light Source, Lawrence Berkeley National Lab, Berkeley, 94720, USA}

\author{Chris Jozwiak}
\affiliation{Advanced Light Source, Lawrence Berkeley National Lab, Berkeley, 94720, USA}

\author{Hadas Soifer}
\affiliation{School of Physics and Astronomy, Faculty of Exact Sciences, Tel Aviv University, Tel Aviv-6997801, Israel}
\affiliation{Center for Light-Matter Interaction, Tel Aviv University, Tel-Aviv 6997801, Israel}

\author{Slavomir Nemsak}
\affiliation{Advanced Light Source, Lawrence Berkeley National Lab, Berkeley, 94720, USA}
\affiliation{Department of Physics and Astronomy, University of California, Davis, CA 95616, USA}

\author{Sergey Y. Savrasov}
\affiliation{Department of Physics and Astronomy, University of California, Davis, CA 95616, USA}

\author{Vsevolod Ivanov}
\email{vivanov@vt.edu}
\affiliation{Virginia Tech National Security Institute, Blacksburg, Virginia 24060, USA}
\affiliation{Department of Physics, Virginia Tech, Blacksburg, Virginia 24061, USA}
\affiliation{Virginia Tech Center for Quantum Information Science and Engineering, Blacksburg, Virginia 24061, USA}

\author{Valentin Taufour}
\affiliation{Department of Physics and Astronomy, University of California, Davis, CA 95616, USA}

\author{Inna M. Vishik}
\email{ivishik@ucdavis.edu}
\affiliation{Department of Physics and Astronomy, University of California, Davis, CA 95616, USA}

\date{\today}

\begin{abstract}	
The multiple crystalline terminations in magnetic Weyl semimetal Co$_3$Sn$_2$S$_2$ display distinct topological and trivial surface states, which have successfully been distinguished experimentally. However, a model of pure terminations is known to be inadequate because these surfaces exhibit a high degree of spatial heterogeneity and point disorder.  Here we perform a spectromicroscopy study of the surface chemistry and surface electronic structure using photoemission measurements in combination with first-principles calculations of core levels.  We identify an intermediate region with properties distinct from both the sulfur and tin terminations, and demonstrate that the spectral features in this region can be associated with a disordered termination with a varying density of surface tin vacancies.  This work establishes heuristics for identifying variable surface disorder using photoemission, an important prerequisite to experimentally establishing the behavior of momentum-space topological surface features subject to variable surface disorder on a single cleave.

%Finally, we show how a combination of algorithmic and machine learning analysis of photoemission data can be used to extract identifying features, classify spatial regions, and correlate local chemistry with local electronic structure.
\end{abstract}
%This work establishes Co$_3$Sn$_2$S$_2$ as a prototype for investigating the fate of Fermi arcs amids variable surface termination on a single cleave
%This work establishes heuristics for identifying variable surface disorder using photoemission, an important prerequisite to experimentally establishing the behavior of momentum-space topological surface features amidst variable surface disorder within a single cleave

\maketitle

\section{Introduction}
Topological quantum materials feature bulk-boundary correspondence, whereby topological properties of the bulk are connected to gapless edge states \cite{Essin:BulkBoundary2011}.  Examples include linearly-dispersing surface bands with spin-momentum locking in topological insulators, hinge modes in higher-order topological insulators, and Fermi arcs in Weyl semimetals.  What is sometimes less appreciated is that the chemical composition of the surface can strongly affect the morphology of characteristic surface states.  This is because different surface terminations can have different polarity, dangling bonds, symmetries, disorder, or orbital content.  This variability can lead to experimental ambiguities \cite{Li_SmB6_mystery2020}, but it can also be used as a tuning knob to drive the presentation of topological surface states within a single material\cite{Yang:FermiArcManipulation_NbAs,Wadge:NbP_decoration, Yue:Bi2Se3_Etch}.

In Weyl semimetals the characteristic Fermi arcs terminate at surface-projections of bulk Weyl points of opposite chirality \cite{YanFelser:WeylReview}. The shape and length of Fermi arcs has implications for surface-dominated transport \cite{RestaFermiArcConductivity2018,Zhang:nanobelts2019}, but bulk topology is only one factor in their morphology. Fermi arc length and connectivity can depend on surface termination, and can be tuned with surface decoration and destroyed with strong disorder \cite{Souma:ObservationOppositeSurfacesNbP2016,Hoffman:PtBi2Termination2025,Wadge:NbP_decoration,li2024disorderdrivennonandersontransition}.  An archetype of variable surface termination yielding different presentations of characteristic boundary states is Co$_3$Sn$_2$S$_2$, a magnetic Weyl semimetal on a Kagome lattice \cite{liu_magnetic_2019,Wang2018,Lohani2023_electron_and_hole_doped,Lohani_2025}. Structurally, it can be viewed as repeating layers of Co$_3$Sn-S-Sn-S.  The Co$_3$Sn termination has nanometer-scale lateral size and is primarily observed by nanoscale probes \cite{Morali2019,Howard:ChiralEdgeCoTerm_2021}; it is predicted to yield short Fermi arcs connecting across Brillouin zones (BZs) \cite{Morali2019}. The Sn and S terminations can have mesoscale extent, measuring up to tens of microns \cite{antonio_2021,mazzola2023}.  The former has the cleanest presentation of long Fermi arcs connecting within the BZ, and most ARPES studies have focused here \cite{liu_magnetic_2019,antonio_2021}.  The distance between the Sn and S layers in the crystal structure is smaller than their bond length, yielding irregular terminations and point defects which have been characterized by scanning tunneling spectroscopy (STS) \cite{Yin2019,jiao_signatures_2019,Liu:PerspectiveStepEdges_2024,Liu:AtomicPerspectiveTopoMagnetism_CSS_2025} but whose effect on mesoscale spectroscopic measurements has not been established.

Here we show spatially-resolved ARPES and XPS data which identify and characterize the disordered region between pure terminations in Co$_3$Sn$_2$S$_2$.  The amount of disorder varies on mesoscale length scales, which we show to be consistent with variable densities of Sn-vacancy defects.  Additionally, we demonstrate how a combination of algorithmic and machine learning analysis of photoemission data can be used to extract identifying features, classify spatial regions, and correlate local chemistry with local electronic structure.

\begin{figure*}[htb]
    \includegraphics[width=0.9\textwidth]{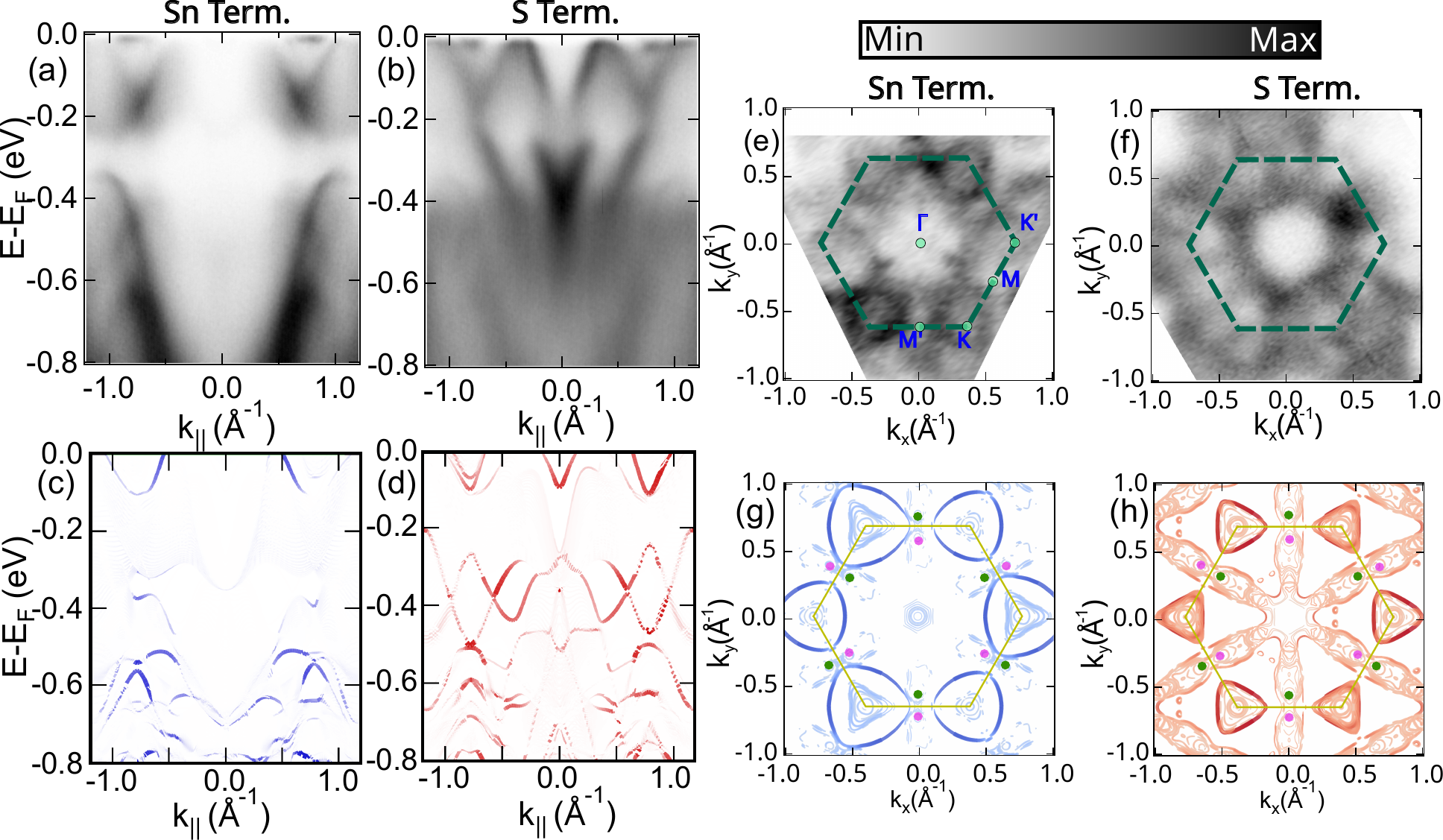}
    \caption{Pure terminations of Co$_3$Sn$_2$S$_2$: ARPES. (a)-(b) ARPES spectra and (c)-(d) corresponding DFT slab calculation along the high-symmetry $\Gamma-K$ direction on the Sn (a,c) and S (b,d) termination, respectively. 
	(e)-(f) constant energy maps at $E_F$ for the Sn and S termination and (g)-(h) corresponding DFT-slab calculation on the Sn (e),(g) and S (f),(h)  terminations, respectively. DFT colors heavily emphasize bands with strong Sn (blue) or S (red) surface character. Weyl points momenta are shown as green and pink dots in panel (g) and (h).}
    \label{heterogeneity}
\end{figure*}

\section{\textbf{Methods}}

Single crystals of Co$_3$Sn$_2$S$_2$ were synthesized using solution growth. A ternary mixture with initial composition Co$_{12}$Sn$_{80}$S$_8$ was first heated to 400\textdegree C over two hours and held for another two hours. It was then heated to 1050\textdegree C over six hours and then held there for 10 hours, followed by slow cooling down to 740\textdegree C within 90 hours. The remaining flux was removed by centrifugation. Shiny hexagonal crystals were obtained. X-Ray powder diffraction data reports a hexagonal cell with $a=5.3641$ \AA \,and $c=13.1724$ \AA \,which are consistent with reported values \cite{Shen_2023}.

Angle resolved photoemission spectroscopy (ARPES) and x-ray photoelectron spectroscopy (XPS) measurements were performed with microfocused synchrotron light at the Advanced Light Source MAESTRO beamline (7.0.2.1) and Beamline 5-2 at the Stanford Synchrotron Radiation Lightsource. All ARPES spectra shown here were obtained with photon energy $115$ eV and a nominal energy resolution of $20$ meV or better. $115$ eV photon energy was previously determined to closely access the Weyl points which are located above the Fermi level ($E_F$) \cite{liu_magnetic_2019}.  S$-2p$ core levels were obtained with photon energy $250$ eV and a nominal resolution better than $100$ meV. Linearly polarized light was used in all measurements. Spectra were collected at a temperature $<30$K, well below T$_c$ \cite{Schnelle:FM_halfMetal_2013}. Samples were cleaved \textit{in-situ} at the measurement temperature. Spatial XPS and ARPES data (Figs. \ref{spatial_maps}-\ref{dft_sessa}) were obtained on a 25$\times$15 grid at MAESTRO.  The beam parameters for those spectra were $18 \mu m \times 19 \mu m$ FWHM, but the analyzer optical axis was $54.75^\circ$ from the beam direction in the horizontal plane, giving a footprint on the sample of $30 \mu m \times 19 \mu m$.  Energy vs momentum ARPES spectra are normalized to intensity above the Fermi level, and linear color scale is normalized to the full range within displayed window.

First-principles calculations were performed using the Vienna ab initio Simulation Package (VASP) \cite{VASP}. Standard hexagonal unit cells containing three Co$_3$Sn$_2$S$_2$ formula units were used to construct a $2 \times 2 \times 5$ unit cell slab containing 420 atoms for the surface state calculations. A vacuum of $50$ \AA \,was introduced to electronically isolate the surfaces. Calculations were performed with no spin orbit coupling and an energy cutoff of $280$ eV on a $3 \times 3 \times 1$ $k$-point grid. Core level energies were computed using the final state approximation; sulfur atoms near the center and surfaces of the cell were selected and an S$-2p$ core electron was removed and placed into the valence band. Additionally, sulfur core level calculations were performed with one or two of the tin atoms removed on the tin-terminated surface. Convergence of the binding energies was tested using unit cells ranging from $2 \times 2 \times 3$ (252 atoms) to $2 \times 2 \times 7$ (588 atoms) \cite{supplement}.

Core level simulations were performed using the Simulation of Electron Spectra for Surface Analysis package (SESSA)\cite{SESSA}. SESSA takes in crystal structure and chemistry information for a given geometry of a photoemission experiment to calculate the relative photoemission core level intensities observed using known electron attenuation lengths and photoemission cross-sections. Due to the dependence of line widths on experimental conditions, we used line widths from the data in this manuscript as inputs for the simulated spectra. Where relevant, DFT was used to calculate chemical shifts in binding energies, which were then used as inputs for simulated spectra. More details of the crystal structure used can be found in the supplements \cite{supplement}.

\section{Results}

\begin{figure}[ht]
    \includegraphics[width=1\columnwidth]{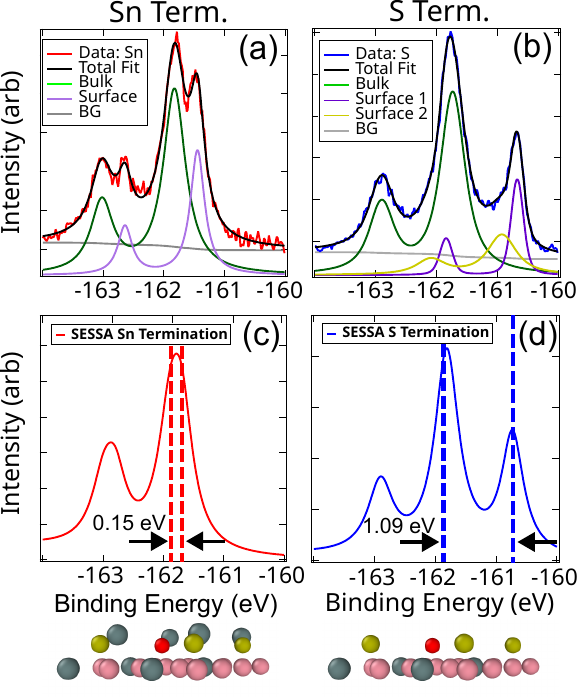}
	\caption{
		S$-2p$ core levels on pure terminations. (a)-(b) S$-2p$ core levels on the Sn- and S- terminations.
        (c)-(d) SESSA simulation of the S$-2p$ XPS spectra at the two terminations using DFT-calculated binding energy shifts and attenuation based on the crystal structure \cite{supplement}.  Binding energy shifts indicated by vertical dashed lines and corresponding labels. Near-surface atomic planes with sulphur (yellow), tin (gray) and cobalt (pink) atoms. A photoemitting sulphur atom is marked red to disambiguate the emitter atom from the surface atom. 
	}
	\label{xps}
\end{figure}

Fig. \ref{heterogeneity} reviews ARPES characteristics of the `pure' Sn and S terminations, as previously established in Refs. \cite{Morali2019}\cite{antonio_2021}\cite{Liu2019}. The two terminations show distinct surface bands. Specifically, Fig. \ref{heterogeneity}(b) (S termination) shows strong features near $\Gamma$ that are absent or minimal in Fig. \ref{heterogeneity}(a) (Sn termination). The surface-like nature of these features on the S termination is supported by DFT slab calculations, shown in Fig. \ref{heterogeneity}(c)-(d). The calculations are colored to exclusively highlight the surface bands; bulk features are discussed in earlier work  \cite{antonio_2021}, and do not differ between terminations.  On the Sn termination, the dominant surface bands near $E_F$ include those that form the Fermi arc surface state.  Viewed as a function of momentum, the Fermi arcs of three adjacent BZs form `triangle-shaped' features at three of the zone corners  \cite{liu_magnetic_2019,antonio_2021,mazzola2023}.   Meanwhile, the S termination shows weaker features at the zone corners, and stronger signatures of the dominant trivial surface state in the interior of the BZ.  It is for these reasons that the Sn termination is the favored one for studying Weyl physics. %The aforementioned characteristics of the Fermi surfaces of each `pure' termination are consistent with DFT (Fig. \ref{heterogeneity}(g)-(h)).

\begin{figure}[b]
    \includegraphics[width=1\columnwidth]{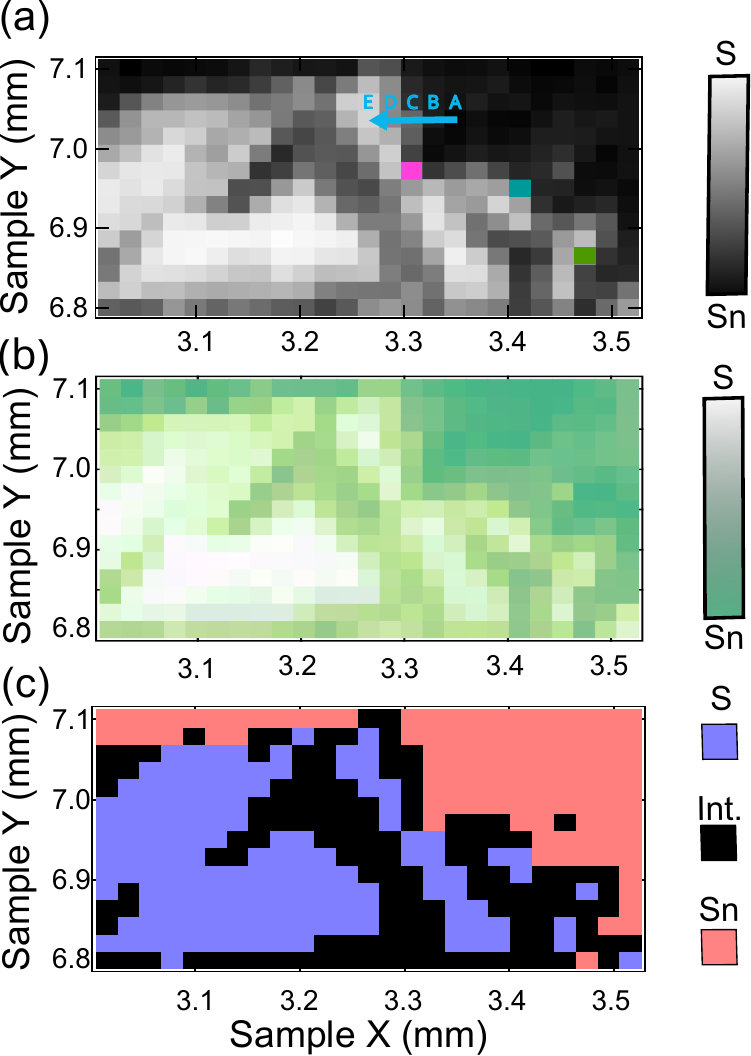}
    \caption{Spatial dependence of S$-2p$ core levels. (a) Spatial map of S$-2p$ S-termination feature, found by integrating in energy window $160.8\pm0.02$ eV at each measurement position. (b) UMAP 3D performed on the same spatial core level data set (c) k-means clustering of the spatial core levels into 3 clusters representing S termination (blue), Tin Termination (Red) and Intermediate (Int.) region (black)}
    \label{spatial_maps}
\end{figure}

Core level photoemission intensity profiles are sensitive to local chemical environments, and the surface sensitivity of UV photoemission allows for detailed characterization of the surface chemistry \cite{briggs_xps1, briggs_xps2}. In Co$_3$Sn$_2$S$_2$, different surface terminations have been shown to manifest in core levels as follows. The Sn$-4d$ levels show an energy shift between terminations that is small or absent \cite{mazzola2023, supplement}.  The Co$-3p$ and Co$-3s$ levels tend to be broad and also show minimal difference between terminations \cite{mazzola2023, supplement}. On the other hand the S$-2p$ core levels show substantial variation in their structure depending on the local environment of the emitter, and are the focus of this work \cite{Liu2019}\cite{Dedkov_2008}\cite{antonio_2021}\cite{ma2024_surfacemod}.  Fig. \ref{xps} summarizes the characteristics of the S$-2p$ core level on the pure terminations. Fig. \ref{xps} (a)-(b)show spectra obtained at two different positions on the same crystal, together with fits. The Sn termination spectra can be fit by two Voigt doublets at different energies, while the S termination requires a third doublet. A Shirley Background, labeled `BG', is used\cite{Major_2020}. The spectra differ primarily via the energy position of the doublet at lower binding energy (purple curves in Fig. \ref{xps}(a)-(b))  \cite{Li_oxidation_2019, antonio_2021, ma2024_surfacemod}. Based on photon-energy dependence experiments, this feature is interpreted to originate closer to the surface \cite{supplement}.  Meanwhile, the doublet at the deepest binding energy (green curves in Fig. \ref{xps}(a)-(b)), is interpreted as originating from bulk sulphur atoms. Binding energy shifts between surface and bulk atoms are commonly observed in core level spectroscopy, and can arise due to both initial state and final state effects in the photoemission process  \cite{Egelhoff_1986}\cite{Eli_93}. To differentiate between these two possibilities, binding energy shifts between bulk and each pure surface termination are computed using DFT-based slab calculations. These calculations yield a surface core level shift of 0.19 eV (Sn termination) and 1.07 eV (S termination) relative to the bulk S$-2p$ electron binding energies, in close agreement with experimental values of 0.33 eV and 1.09 eV respectively. The difference in binding energy between bulk and surface can, thus, largely be attributed to initial state effects from the unsatisfied valences at the surface. The deviations from experimental values may occur because of the finite cluster size of the calculations, core hole dynamics, and other final state effects that are neglected. Fig. \ref{xps} (c)-(d) shows SESSA simulated S$-2p$ core level spectra, where core-level shifts of the S atoms closest to each surface are inputted from DFT.
%add: shifting of core levels is controversial

\begin{figure}[ht]
    \includegraphics[width=1\columnwidth]{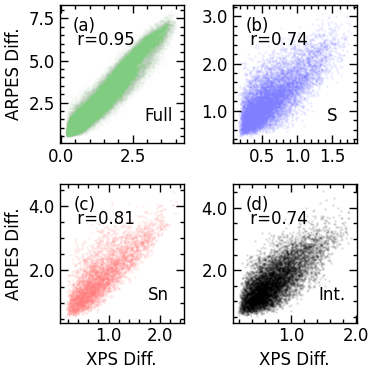}
    \caption{Correlation between local S$-2p$ core levels and ARPES spectra. Scatter plots of XPS and ARPES euclidean difference metrics for the full data set (a), as well as S (b), Sn (c), and mixed (d) cluster subsets. $r$ is the Pearson correlation coefficient.}
    \label{cluster}
\end{figure}

Thus far, we have only shown spectra at single positions identified as pure terminations, but photoemission spectroscopies with small spot size can also be employed as spectromicroscopies to ascertain mesoscale electronic phenomena  \cite{Hideaki_2018}\cite{Horacio_2015}. Fig. \ref{spatial_maps} examines the S$-2p$ core level heterogeneity across a $0.5$ mm $\times 0.3$ mm area grid on the sample using three different visualization schemes. Fig. \ref{spatial_maps} (a) shows a spatial map that is produced by integrating the S$-2p$ core level intensity at a binding energy of $160.8\pm0.02$ eV, to highlight the spatial variation of the spectral feature previously identified with surface component of the S-termination spectrum. The highest intensity regions on the map are interpreted as the S termination and the darkest regions as the Sn termination. Fig. \ref{spatial_maps}(b) employs uniform manifold approximation and projection (UMAP) to project the high-dimensional photoemission spectra onto lower dimensional spaces, while preserving its topological structure \cite{umap}. Here, high-dimensional refers to the $800\times1000$ pixels in an ARPES spectrum or the $1000$ pixels of an XPS data array. 
%This becomes useful for highlighting subtle differences in spectroscopic data, as will be seen for example, in different spectra within the intermediate region in Fig.\ref{spatial_arpes}.
In Fig. \ref{spatial_maps}(b), we apply UMAP to map every point on the spatial grid of S$-2p$ XPS spectra onto a 3D space. The resulting coordinates for each point are then mapped to a color scale. The generated color map closely resembles the intensity plot obtained in Fig. \ref{spatial_maps}(a), but without use of user input.  Because this method appears to capture the same spatial distributions as Fig. \ref{spatial_maps}(a), we identify the dark green as `Sn termination' and the white as `S termination' on the color scale. As another way to characterize the different surface features, the spatial core level data were grouped into three clusters using k-means clustering  \cite{Iwasawa:Unsupervised_2022,Imamura_kmeans_ARPES_2024, Majchrzak_2025} (Fig. \ref{spatial_maps}(c)). These three clusters correspond closely to the regions in Fig. \ref{spatial_maps}(a)-(b) identified as the S and Sn terminations, plus something in between. The latter is thereafter called the `intermediate region'. %This method yields similar results to the previous two methods but is much less rich in the information it conveys. It also requires an understanding of the surface in the input of the number of clusters. 

%The results shown in earlier figures relies on the experimenter to identify distinguishing spectral features in pure and intermediate termination regions on the surface of Co$_3$Sn$_2$S$_2$.  For this material, this effort is aided by prior work, and by the fact that the S-2p core levels and high-symmetry ARPES cuts yield distinguishing features on the S termination with minimal energy and momentum overlap with features on the Sn termination. Other materials may not yield such clear characteristics, and for that reason, we use this dataset to validate other approaches to segmenting spatially-resolved and multimodal photoemission data.
%say more about clusters

\begin{figure}[hb]
	%\captionsetup{justification=raggedright}
	\includegraphics[width=\columnwidth]{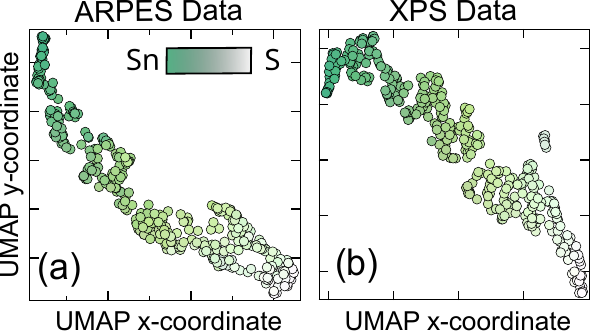}
	\caption{
        UMAP on spatially resolved photoemission data. (a) 2D UMAP on spatial ARPES data. (b) 2D UMAP on spatial S-2p spectra measured at the same spatial positions. 
	}
	\label{UMAP}
\end{figure}

XPS and ARPES provide two complementary information streams on the same cleave, and if the two are correlated, it can be used to understand variations in band structure features in terms of local chemistry. Fig. \ref{cluster} examines and correlates spatial ARPES and spatial XPS data collected on the same grid. This analysis is done for the entire data set (Fig. \ref{cluster}(a)) and for each termination type individually (Fig. \ref{cluster}(b)-(d)), and details are discussed in supplementary materials \cite{supplement}.  For the latter,  groupings from Fig. \ref{spatial_maps}(c) were used.  The degree of correlation between ARPES $\Gamma-K$ cuts and S-$2p$ core levels is quantified using a Pearson correlation coefficient (r).  A coefficient close to one indicates that data points exist on a tight diagonal where a small difference between two XPS spectra always implies a small difference between ARPES spectra and vice versa. The calculated Pearson coefficients for the full dataset, S, intermediate, and Sn termination are 0.95, 0.74, 0.74 and 0.81 respectively.  These high degrees of correlation, especially for the full data set, indicate that local S-$2p$ spectra can be used to select accompanying characteristic ARPES spectra in this compound; this is not always the case, as discussed in supplementary materials \cite{supplement}. 

 Fig. \ref{UMAP} (a) and (b) further establishes spatial evolution of spectra via UMAP projection of both the spatial ARPES and core level data onto a 2D space. The individual data points are color coded to match Fig. \ref{spatial_maps}(b). This visualization is intended to highlight relationships between data points, but the x- and y-axes are not necessarily physically meaningful. The axis coordinates in UMAP are affected by the stochastic nature of the algorithm, thus, qualitative features are more reliable than absolute metrics in the resulting projection. This implies that while general relationships can be ascertained through UMAP, the relative distances of data points can be misleading.
 
 Both of the 2D representations in Fig. \ref{UMAP}(a)-(b) appear as a nearly one-dimensional meander: points at either end correspond to pure terminations, and there is a gradual evolution in between. This indicates a continuous transition between the two terminations with contributions from multiple intermediate configurations between the two terminations. Two ways that a continuous evolution of spectra can arise is 1) if all intermediate spectra are linear combinations of the two endpoints or 2) if spectral features evolve continuously between two endpoints but are not linear combinations of the two endpoints.  Scenario (1) would imply that intermediate points are an average of two pure terminations, while scenario (2) would imply that there are unique characteristics of intermediate spectra.  Disambiguating between the two requires closer examination of the data.

 %a linear combination means that any intermediate state is just the average of the two terminations, whereas option (2) means there are some unique properties that are related to the disorder

\begin{figure}[hb]
	%\captionsetup{justification=raggedright}
	\includegraphics[width=1\columnwidth]{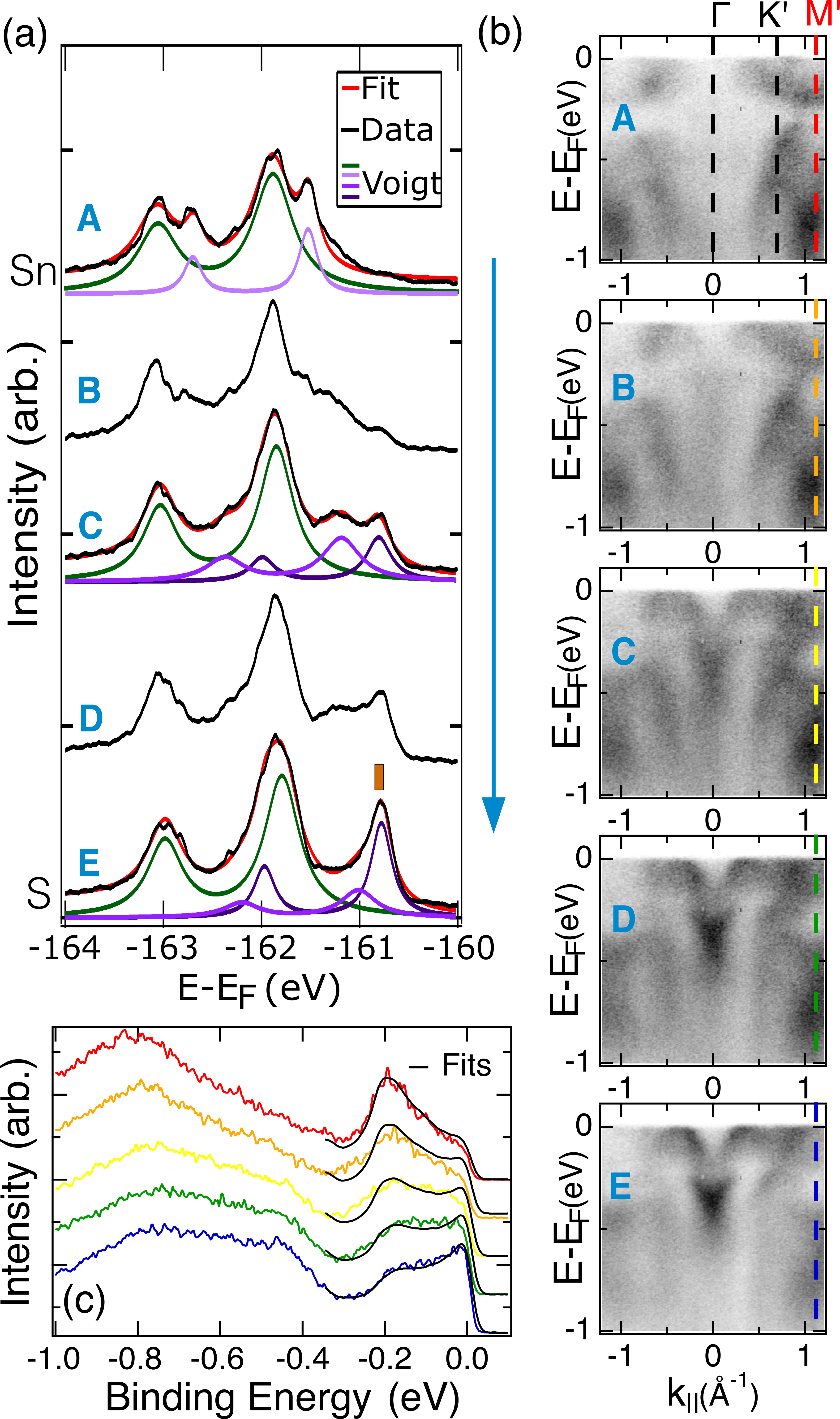}
	\caption{
        Evolution of photoemission spectra between two pure terminations. (a) XPS spectra evolution from Sn to S termination; traces A-E correspond to spatial locations in line cut denoted by arrow in Fig. \ref{spatial_maps}(a) and go from Sn termination (A) to Sn termination (E) across three intermediate spectra.(b) ARPES spectra along $\Gamma-K'-M'$ cut for the same spatial positions in a (c) EDCs at the $M'$ point for each of the spectra in panel (b). Solid black lines are best fits of linear-combinations of spectra from Sn and S termination. %S Trace percentages of $6.3\%$ (Red), $27.0\%$ (Orange), $49.4\%$ (Yellow), $69.0\%$ (Green) and $87.9\%$ (Blue) were obtained for the best fits
	}
	\label{spatial_arpes}
\end{figure}

 Fig. \ref{spatial_arpes} shows the typical evolution of the photoemission spectra for five adjacent positions traversing a termination change (blue arrow in Fig. \ref{spatial_maps}(a)). The S-$2p$ core level is shown along the indicated trajectory in (a). Fits are performed for three selected spectra, labeled `A',`C' and `E'.  A and E represent the pure Sn and S terminations, respectively, while C is representative of the 'intermediate' termination.  Importantly, the fit for spectrum C includes an additional doublet with binding energy distinct from the surface-identified doublets observed for the two pure terminations.  There is no way to produce spectrum C from a linear combination of the pure-termination spectra (A and E), which motivates interpreting the new doublet as originating from a distinct chemical environment of the emitting S atom.  This additional peak is also reminiscent of the third doublet (yellow) in Fig. \ref{xps}(b), albeit with larger relative intensity. The broadness of the additional doublet suggests either the presence of multiple underlying peaks and/or structural disorder. 
 
\begin{figure}[ht]
    \includegraphics[width=0.9\columnwidth]{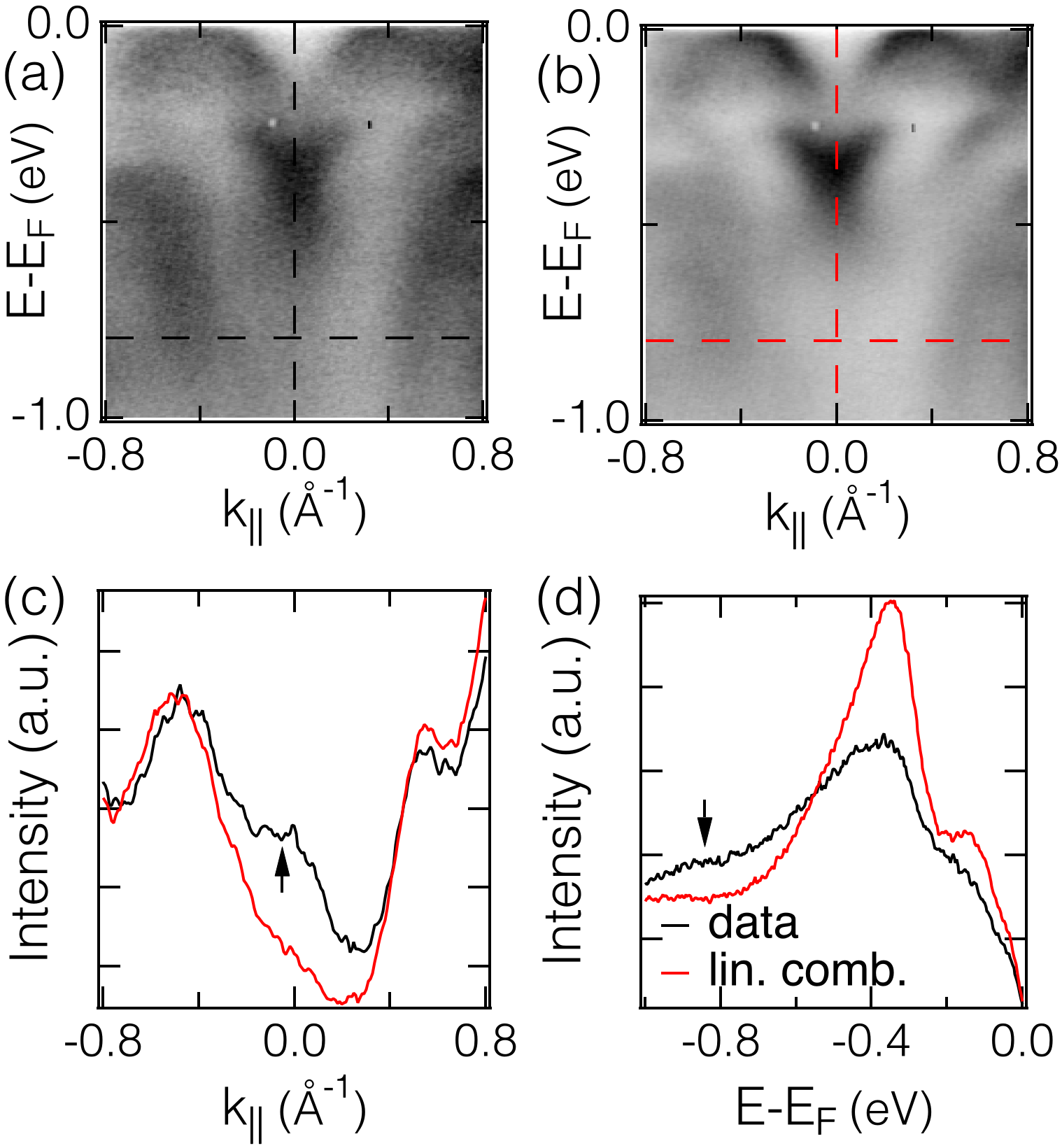}
    \caption{Intermediate region ARPES spectra and deviation from linear combination of pure terminations. (a) Measured spectrum obtained from averaging five points in the XPS UMAP with nearly identical UMAP coordinates. (b) Reconstructed spectrum using a linear combination of 49.8\% S termination and 50.2\% Sn termination. (c) MDCs and (d) EDCs taken along the dashed line positions in (b), showing a comparison between the measured (a) and the reconstructed spectrum (b).}
    \label{difference}
\end{figure}

 We compare this with the evolution of ARPES spectra for the same five positions.  The spectra shown in Fig. \ref{spatial_arpes}(b) reveal a gradual gain of spectral intensity near $\Gamma$ when transitioning from the Sn to the S termination \cite{supplement}. Meanwhile, at the $M'$ point (vertical dashed lines in Fig. \ref{spatial_arpes}(b)), there is a gradual shift of spectral weight to deeper binding energies, which is quantified in the energy distribution curves (EDC) shown in Fig. \ref{spatial_arpes} (c).  In the intermediate regions, signatures of both endpoint terminations coexist. Fig. \ref{spatial_arpes} (c) demonstrates this by fitting the EDCs at the $M'$ point in the intermediate region with linear combinations of spectra from the two terminations \cite{supplement}. The percentage of S traces in the resulting fits are $6.3\%$ (Red), $27.0\%$ (Orange), $49.4\%$ (Yellow), $69.0\%$ (Green) and $87.9\%$ (Blue) for the respective EDCs. The $M'$ point was chosen for this analysis because bulk and surface bands are at an extremum of energy \cite{antonio_2021}. 
 However, we note that ARPES spectra in the intermediate region are not perfect linear combinations of the pure endpoints, which is highlighted in Fig. \ref{difference} and supplementary materials \cite{supplement}.  Fig. \ref{difference} focuses closer to the $\Gamma$ point, and compares a measured spectrum (Fig. \ref{difference}(a)) in the intermediate region with a spectrum reconstructed from a linear combination of pure endpoints (Fig. \ref{difference}(b)).  To enhance statistics, the measured spectrum is an average of five spectra at five different positions, selected from similar XPS UMAP coordinates (Fig. \ref{UMAP}(a)).  This selection procedure is justified by the high degree of correspondence between XPS and ARPES spectra demonstrated in Fig. \ref{cluster}.  The chosen spectra with similar UMAP coordinate also had a similar admixture of Sn and S spectral character at the $M'$ point, and the average value was used to construct Fig. \ref{difference}(b) with a spectral composition of 50.2\% S termination and 49.8\% Sn termination.  The `measured' (panel a) and `linear combination' (panel b) ARPES spectra differ in that the former is broader and has more spectral weight at deeper binding energy and normal emission creating a `vertical feature.'  Characteristic momentum distribution curves (MDCs) and EDCs are compared to further illustrate these differences.  Fig.\ref{difference}(c) compared MDCs for data and linear combination (reconstructed) spectra taken at $E_B=-0.8 eV$.  An arrow highlights extra peak in data corresponding to the `vertical feature.'  This feature is also seen in EDCs (Fig. \ref{difference}(d)), which further demonstrate spectral broadness of intermediate region ARPES spectra.  %Common features appear in data and reconstructed spectra, but they are broader in the former and there is additional spectral weight at higher binding energy corresponding to the vertical dispersion feature.

 %and several examples of the deviations are highlighted in Fig. \ref{spatial_arpes} and supplements.  These differences include: 1) enhanced spectral weight at $\approx -0.3 eV$ in the $M^\prime$ EDC 2)    % However, we additionally note that ARPES spectra in the intermediate region (B-D) show enhanced  spectral weight at deeper binding energy at $\Gamma$ \cite{supplement}, contributing to a `vertical dispersion' feature, a motif that appears in systems with strong disorder  \cite{Ciocys2024}. 

\begin{figure*}[htb]
%	\captionsetup{justification=raggedright}
    \includegraphics[width=0.9\textwidth]{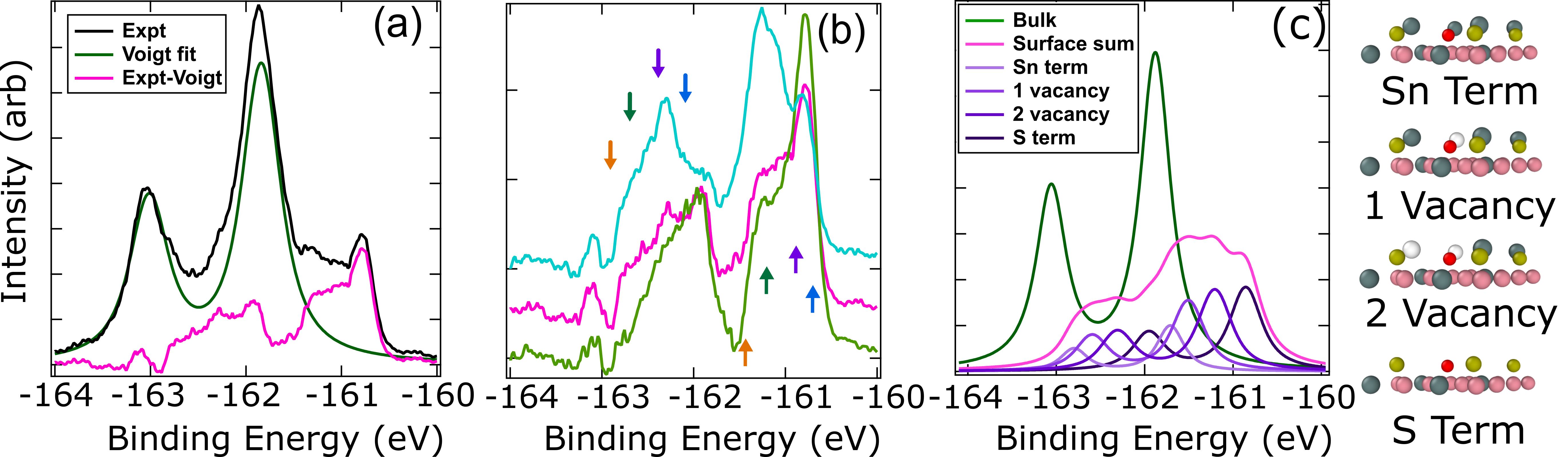}
	\caption{
        Intermediate region S-$2p$ surface spectra and simulations. (a) Spectrum at pixel marked in pink in Fig. \ref{spatial_maps}(a), illustrating subtraction of bulk S$-2p$ peak. (b) Example spectra following subtraction procedure in (a), with several distinct residual surface peaks highlighted by arrows. Colors correspond to spatial locations of the colored pixels in Fig. \ref{spatial_maps}(a). (c) SESSA simulation of surface spectrum with an even coverage of four spectra indicated on right, using DFT-input binding energy shifts.
	}
	\label{dft_sessa}
\end{figure*}

 %local vs delocalized electrons
To examine the origin of the spectroscopic features of the intermediate region further, we select S-$2p$ spectra from several other example spatial locations within the intermediate region, labeled with pink, dark blue, and green pixels in Fig. \ref{spatial_maps} (a). These three positions are chosen to illustrate spectral variety in the intermediate region. The surface contribution to these spectra can be mostly isolated by subtracting the dominant doublet associated with the bulk (green).  An example of this is illustrated (pink) in Fig. \ref{dft_sessa} (a) for data from the pink pixel in Fig. \ref{spatial_maps}(a). Repeating this subtraction procedure at other positions, we find that the spectral intensity differences are peaked at multiple binding energies across the intermediate region, highlighted by arrows at approximately $\sim162.6$ eV, $\sim162.4$ eV, $\sim162.1$ eV and $\sim161.9$ eV, as well as their spin-orbit-split pair (Fig. \ref{dft_sessa} (b)). It is noted that the small peak at $\sim163.1$ eV and its pair at $\sim161.8$ eV are artifacts of incomplete bulk peak subtraction, and are thus not included in the labeling of various near-surface S-$2p$ peaks. These multiple binding energies of near-surface S-$2p$ states, Fig. \ref{dft_sessa} indicates multiple chemical environments of near-surface sulphur atoms. We capture the binding energy shifts of the near-surface S-$2p$ core levels by considering different number of Sn vacancies above the sulphur layer using DFT. The calculations show increasing binding energy of the surface S$-2p$ shoulder and the transformation of XPS line shape for an increasing number of Sn vacancies above a near-surface sulphur atom \cite{supplement}. A photoemission experiment with a mesoscale spot size sees all of these spectral types superposed at each position. The simultaneous presence of these features along with the S termination shoulder in the experimental data presented in Fig. \ref{spatial_arpes}, in Fig. \ref{xps} (b) and in \ref{dft_sessa}(b) can be attributed to the macroscopic nature of our measurements compared to the nanoscale origin of these defects. To further model observed spectra in Fig. \ref{dft_sessa}(b), the calculated binding energy shifts are used as inputs for SESSA simulations of XPS spectra for just the surface. Equal compositions of all the distinct surface states are used as an example. Qualitative agreement is shown; we note that the varying lineshapes of subtracted spectra in Fig. \ref{dft_sessa}(c) indicates mesoscale variation in concentration of each defect type.  %

  %Indeed, the spectra in Fig. \ref{dft_sessa}(b) contain at least four distinct binding energies with different relative weights at different positions. 

  %This suggests a model of the intermediate S$-2p$ states as Sn vacancy states, resulting in partial valency fulfillment on the photoemitting sulphur atom. This partial valency causes the intermediate binding energy, similar to the valency difference causing the surface core level shift. 
  %took this out because same idea is in discussion

\section{Discussion}
%The presence of multiple terminations in Co$_3$Sn$_2$S$_2$ has been previously observed, through both photoemission and scanning tunneling microscopy studies\cite{Morali2019}. In particular, heterogeneity in photoemission has also been observed \cite{mazzola2023}. However, a systematic study correlating the surface chemistry to the observed bands is lacking in the literature. A paradigm to study surfaces more generally while performing photoemission measurements also needs to be established. Chemical and electronic differences can be studied systematically and simultaneously using a combination of ARPES and XPS, but large-scale spatial heterogeneity studies require the employment of machine-learning models to aid analysis. This work aims to provide the tools through a systematic study of the surface and the implementation of topological machine learning in a system with surface features of scientific interest.

%As captured in the results section, 
%As previously discussed, Fig. \ref{heterogeneity} captures the ideal extremes of the two dominant terminations in the material from a band structure perspective. We observe remarkable, yet predictable, differences in the surface states at the two terminations. In addition, as observed in Figure \ref{xps}, the high quality of the spectra are correlated to lower presence of disorder. 
First, we discuss the different spectromicroscopy data processing methods employed in this manuscript, and how they contributed to the identification of the intermediate region. Then, we turn to the interpretation and significance of the intermediate region.

The most basic method for identifying surface termination relies on manual identification often followed by correspondence with DFT. This is adequate for systems that have been studied before, systems that have very clear distinction between spectra on different terminations, materials where DFT has good correspondence with ARPES spectra, and materials where a model of predominantly pure terminations is accurate.  Co$_3$Sn$_2$S$_2$ is an example of a system where a large portion of the cleaved surface shows spectra that are not pure terminations, and instead shows a continuum of spectra.  Additionally, Co$_3$Sn$_2$S$_2$ shows only qualitative correspondence between ARPES spectra and DFT slab calculations: the preponderance of trivial surface states on the S termination and Fermi arcs that are sensitive to time-reversal symmetry breaking below $T_c$ on the Sn termination are captured, but exact binding energies of features are not \cite{antonio_2021,mazzola2023,Liu:TopoPT_2021}.  It should be noted that the correspondence between experiment and DFT for the characteristic S$-2p$ core levels in this system calculated here is more exact (Figs. \ref{xps}, \ref{dft_sessa}).  

Following manual identification of distinguishing spectral features, the spatial extent of each termination can be quantified by integrating over a chosen window in each photoemission spectrum in a spatial map.  This has the advantage of being easy to implement, but the disadvantage of inadequacy with overlapping features and not being able to distinguish when there is low intensity due to the absence of a feature or due to a noisy spectrum.  

The next level of spatial data grouping in this manuscript is k-means clustering, an unsupervised machine learning method to separate \textit{m} data points into \textit{k} clusters based on a sense of distance from the center of each cluster \cite{Melton_2020,Iwasawa:Unsupervised_2022,Ekahana_2023,Imamura_kmeans_ARPES_2024}.  This has the advantage of not relying on manual identification, but the disadvantage of requiring a user-chosen number of clusters. It is ideal for systems where there are a known/fixed number of spectral types, and less ideal otherwise.  

In this manuscript, we contrast these methods with UMAP projected to three dimensions, which does not use experimenter-identified distinguishing spectral features or number of clusters. UMAP has emerged as a robust tool for dimensionality reduction because unlike algorithms such as principal component analysis (PCA) that perform linear dimensional reduction, UMAP leverages non-linear manifold learning to accurately capture both local neighborhoods and the broader global structure of the data \cite{umap_rev}.  Interestingly, a map produced in this manner is nearly indistinguishable from one produced by integrating the distinguishing portion of the spectra (Fig. \ref{spatial_maps}(a)-(b)).  As such, at least for Co$_3$Sn$_2$S$_2$, the more agnostic UMAP grouping appears to capture the salient features of the spatial distribution of different spectral types.  Additionally, the UMAP analysis captures the smooth evolution between two endpoint spectra (pure terminations), both in the ARPES and XPS channel, which was instrumental in identifying the intermediate region. %We note that UMAP has had minimal application to ARPES data thus far.

Prior work has discussed correspondence between DFT slab calculations and ARPES spectra on pure terminations in Co$_3$Sn$_2$S$_2$ \cite{mazzola2023}\cite{antonio_2021}\cite{Lohani_2025}. Here, we extend these works by including DFT calculations of surface core level chemical shifts and considering the disordered intermediate region.  When considering pure terminations (Fig. \ref{xps}), the agreement between core-level DFT calculations and experiment suggests the following model for the surface core level binding energy shift. The S and Sn layers in the bulk crystal structure consist of alternating positive and negative oxidation states, with binding energies consistent with S$^{2-}$ and Sn$^{2+}$ \cite{Liu2018}. Upon cleaving on a S termination, this surface consists of S atoms with unsatisfied valence. At the Sn surface, the S$-2p$ electrons are attracted to the Sn atoms above them, in contrast to the S termination. This produces a binding energy difference for the S$-2p$ electrons between the two terminations. Extending the aforementioned model to a Sn surface with missing atoms would imply that the S atoms surrounding these vacancies would have binding energy shifts intermediate between those of the pristine terminations. 

Fig. \ref{dft_sessa}(c) confirms this intuitive model, demonstrating several discrete intermediate energy S$-2p$ binding energies with intensities that vary between spatial points,  presumed to depend on vacancy concentrations based on results of calculations. The presence of Sn vacancies has previously been reported on the UHV-cleaved surface of the material through large area STS studies \cite{Yin2019,Gao2021_defects}, but their signature with the present mesoscale photoemission probe suggests that this constitutes a sizable fraction of the cleaved surface. We also note that step-edges are another prominent structural feature reported by STM \cite{Liu:PerspectiveStepEdges_2024,Liu:AtomicPerspective_2025}; by the qualitative model discussed above, these should give similar signatures in XPS with mesoscale beam spot, and cannot be distinguished with present measurements.  We note that the presence of an extended (mesoscale) disordered region does not appear ubiquitously in all compounds with variable surface termination in the literature \cite{Iwasawa:BuriedCuOChains_2019,Nakayama:Nanomosaic_2019,Jung_2022_LaFeAsO}, though this topic has been a lesser focus and merits further investigation in compounds where surface disorder is an important variable. In the present compound, because of the high degree of correlation between XPS and ARPES (Fig. \ref{cluster}), as well as the smooth evolution between the two pure terminations (Figs. \ref{UMAP},\ref{spatial_arpes}), we can select ARPES spectra corresponding to different surface disorder environments which in turn show ARPES features of disorder (Fig. \ref{difference} and Ref. \cite{supplement}) to different degrees.  %add references

Spatially-resolved ARPES spectra in Co$_3$Sn$_2$S$_2$ in the intermediate region \textit{cannot be fully described} as a linear combination of the pure terminations, and deviations from a simple linear-combination picture are important demonstrations of spectral features arising from a chemically disordered surface. The most prominent ARPES feature found distinctly and exclusively in the intermediate region is a `vertical' feature at normal emission.  A `vertical dispersion' feature has been reported in ARPES spectra on systems with strong disorder including amorphous compounds  \cite{Ciocys2024}, correlated electron systems \cite{Meevasana:Heirarchy2007,Chang:DispersionKinksFeSC_2024,Sun:ElectronicStructureNickelates2025}, and surfaces following in-situ deposition of impurities/dopants \cite{Ryu:BlackPhosphorus_surface_dopant_2020}. Unlike the aforementioned examples, here we demonstrate such a spectral feature to arise from naturally occurring surface defects.  It serves as a signature of surface disorder, in addition to the characteristic S-$2p$ core levels in the intermediate region (Figs. \ref{spatial_arpes},\ref{dft_sessa}) and measured ARPES spectra that are broader than linear combinations of pure terminations (Fig. \ref{difference}).  %Additional features of the disordered intermediate region includes overall broader spectra (Fig. \ref{difference}(d)) as well as the region near -0.3 eV at the M point binding energy (Fig. \ ref{spatial_arpes}(c)) where intermediate-region spectra consistently exceed intensity of either pure termination.

Surface disorder is an important motif in driving novel behaviors in quantum materials \cite{Schubert:fateOfTISurfaceStateStrongDisorder_2012,Peng:SpinCurrentSurfaceDisorderTI_2016,Slager:DissolutionTopoFermiArcs_2017,Wilson:DoFermiArcsSurviveDisorder_2018}. Even though the disorder itself is often nanoscale, the relevant spectroscopic or transport effects can be meso/macro scale. In Weyl semimetals, the fate of Fermi arcs amid surface disorder appears to depend on disorder strength \cite{Sessi:ImpurityScreeningStabilityFermiArcs2017,Slager:DissolutionTopologicalFermiArcs_2017,Wilson:FermiArcDisorder2018,Brillaux:FermiArcDisorderStrength2021} and disorder may also drive quantum phase transitions in the near-surface region \cite{li2024disorderdrivennonandersontransition}.  Notably, Co$_3$Sn$_2$S$_2$ can yield \textit{multiple disorder environments within a single cleave}, which can be identified successfully from S-$2p$ core levels using the methodology in this manuscript.  This methodology can also be used to identify the variable spectroscopic landscape using ARPES alone, for example, in low-photon energy experiments that do not have access to core levels. The spectral variation of the intermediate region is evidenced by the variability in the surface contributions to the S-$2p$ spectra (Fig. \ref{spatial_arpes}(a),\ref{dft_sessa}(b)), which is understood via variations of densities of different types of defects at different mesoscale positions. It is also evidenced by the smooth evolution of both ARPES and XPS in the UMAP dimensionality reduction (Fig. \ref{UMAP}), as opposed to several discrete clusters. Additionally, the quantification of pairwise spectral differences in Fig. \ref{cluster} indicates that there are different types of intermediate S-2p spectra, which correlate well with accompanying ARPES spectra.  These heuristics can be used in future work exploring the evolution of Fermi arcs in the disordered intermediate region of Co$_3$Sn$_2$S$_2$, or they can be applied to other materials where a variable disorder environment can be achieved on a single surface.
%add refs to this paragraph
%add: other phenomena that depend on surface disorder
%add: non triviality of mososcale intermediate region

\section{Conclusions}
Co$_3$Sn$_2$S$_2$ has established itself as an important standard candle for implications of variable cleave terminations, because the characteristic Fermi arcs present distinctively on different surfaces. Prior ARPES literature has focused on identifying and measuring the \textit{pure} terminations.  In this paper, we delved into the less-discussed and highly disordered intermediate region, typically found between pure terminations, shedding light on how local defect environments produce characteristic S-$2p$ surface states with discrete binding energies different from pure terminations. These characteristic local S-$2p$ spectra correspond closely with specific ARPES spectra with distinct spectroscopic features of disorder, making Co$_3$Sn$_2$S$_2$ a fruitful platform for investigating the effect of variable disorder on topological surface states. Additionally, we discussed different methods for finding structure within multidimensional photoemission spectromicroscopy data, noting the propensity of UMAP to highlight small but relevant spectroscopic differences and identify relationships within data without user input.\\

\begin{acknowledgments}

We acknowledge helpful discussions with Donghui Lu, Makoto Hashimoto, Journey Byland, Yucheng Guo, Maximilian Jaugstetter, Lorenz Falling, Jay Hineman, Alex Hexemer,Wiebke K\"opp, and Tanny Chavez. The ARPES and XPS measurements in this work were supported by the US-Israel Binational Science Foundation Grant 2020067. Analysis of intermediate region data by MKC was supported by NSF-DMR 2428464.  Use of the Stanford Synchrotron Radiation Lightsource, SLAC National Accelerator Laboratory, is supported by the U.S. Department of Energy, Office of Science, Office of Basic Energy Sciences under Contract No. DE-AC02-76SF00515. This research used resources of the Advanced Light Source, a U.S. DOE Office of Science User Facility under contract no. DE-AC02-05CH11231.  SS and MS were supported in part by an ALS Doctoral Fellowship in Residence. VI acknowledges support from the National Science Foundation Growing Convergence Research Award 2428507. S.S acknowledges support from US DOE grant No. DE-SC0026106. Z.S. and V.T. acknowledge the Cahill research fund for supporting sample synthesis. H.S. acknowledges the support of the Zuckerman STEM Leadership Program, the Young Faculty Award from the National Quantum Science and Technology program of the Israeli Planning and Budgeting Committee. I.S and H.S acknowledge support of ERC project PhotoTopoCurrent 101078232.
%Add Valentin funding acknowledgement
\end{acknowledgments}
\begin{center}\textbf{DATA AVAILABILITY}\end{center}
The data that support the findings of this manuscript will be made openly available following article acceptance.

\bibliography{references}

\end{document}

\setlength{\intextsep}{5pt}
\title{\Large
Supplementary Material: Mesoscale variations of chemical and electronic landscape on the surface of Weyl semimetal \texorpdfstring{Co$_3$Sn$_2$S$_2$}{Co3Sn2S2} visualized by ARPES and XPS
}
\author[1]{Sudheer Anand Sreedhar}

\author[1]{Matthew Staab}

\author[1]{Mingkun Chen}

\author[1]{Robert Prater}

\author[1]{Zihao Shen}

\author[2]{Giuseppina Conti}

\author[3,4]{Ittai Sidilkover}

\author[5]{Zhenghong Wu}

\author[2]{Eli Rotenberg}

\author[2]{Aaron Bostwick}

\author[2]{Chris Jozwiak}

\author[3,4]{Hadas Soifer}

\author[1,2]{Slavomir Nemsak}

\author[1]{Sergey Y. Savrasov}

\author[6,7,8]{Vsevolod Ivanov}

\author[1]{Valentin Taufour}

\author[1]{Inna M. Vishik}

\affil[1]{Department of Physics and Astronomy, University of California, Davis, CA 95616, USA}
\affil[2]{Advanced Light Source, Lawrence Berkeley National Lab, Berkeley, 94720, USA}
\affil[3]{School of Physics and Astronomy, Faculty of Exact Sciences, Tel Aviv University, Tel Aviv-6997801, Israel}
\affil[4]{Center for Light-Matter Interaction, Tel Aviv University, Tel-Aviv 6997801, Israel}
\affil[5]{Department of Physics, Carnegie Mellon University, Pittsburgh, PA, 15213, USA}
\affil[6]{Virginia Tech National Security Institute, Blacksburg, Virginia 24060, USA}
\affil[7]{Department of Physics, Virginia Tech, Blacksburg, Virginia 24061, USA}
\affil[8]{Virginia Tech Center for Quantum Information Science and Engineering, Blacksburg, Virginia 24061, USA}

\date{}
\maketitle
\vspace{-10ex}

\setcounter{secnumdepth}{0}
%%%%%%%%%%%%%%%%%%%%
% Slab Calcs & DLM %           
%%%%%%%%%%%%%%%%%%%%

\section{\large Supplementary Note 1: Convergence of core level calculations.}

\begin{table}[h]
    \centering
    \begin{tabular}{ c | c | c c }
        \hline
        \hline
        Slab Size & \# of atoms & Binding energy, S-termination (eV) & Binding energy, Sn-termination (eV) \\\hline
        %Compression & \multicolumn{2}{c|}{ $U = 4.0$eV} & \multicolumn{2}{c}{ $U = 5.0$eV} \\
        $2 \times 2 \times 7$ & 588 & 68.54024655 & 67.6419138  \\
        $2 \times 2 \times 6$ & 504 & 68.53135813 & 67.63895787 \\
        $2 \times 2 \times 5$ & 420 & 68.54031705 & 67.65324571 \\
        $2 \times 2 \times 4$ & 336 & 68.54417584 & 67.63852258 \\
        $2 \times 2 \times 3$ & 252 & 68.5406361  & 67.65256036 \\
        $2 \times 3 \times 4$ & 504 & 68.53947566 & 67.63363181 \\
        \hline
        \hline
    \end{tabular}
    \caption{S$-2p$ core level binding energies for a sulfur atom at the S surface termination and Sn surface termination for different slab sizes. The vacuum in all cases is 50 \AA. }
    \label{dft-core-levels}
\end{table}

\section{\large Supplementary Note 2: Determination of Bulk and surface features}
\begin{figure}
    \centering
    \includegraphics[width=0.5\linewidth]{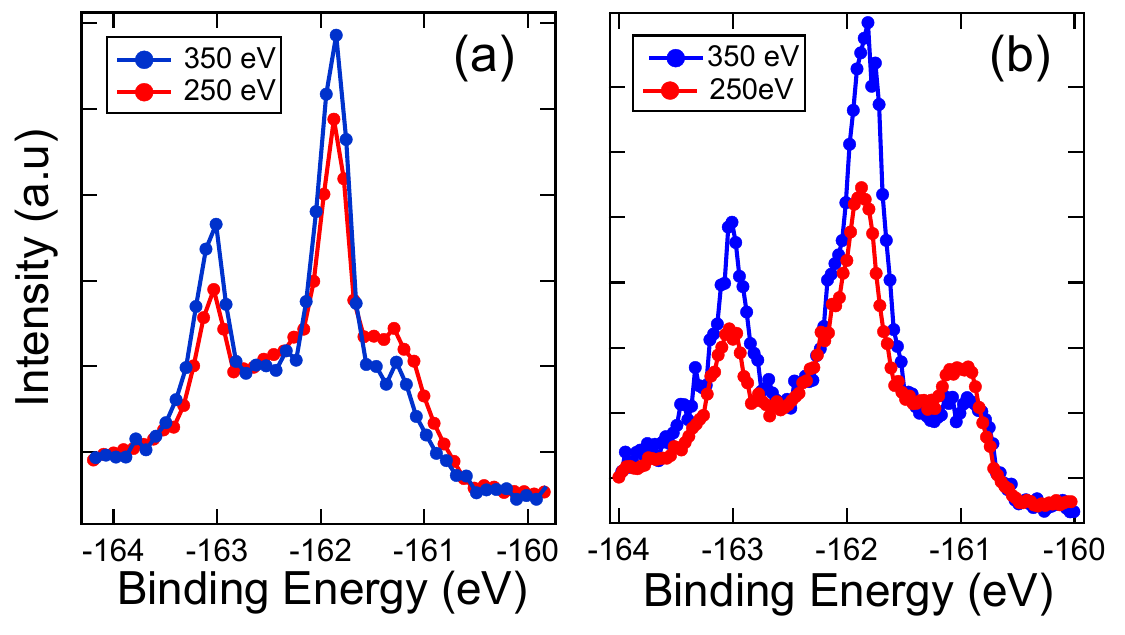}
    \caption{Photon Energy dependence of the S$-2p$ core level in (a) Sn Termination (b) S Termination. Core levels have been normalized for photon flux and photoionization cross-sections}
    \label{supp_pe}
\end{figure}
Fig \ref{supp_pe} shows Photon-flux and cross-section normalized S$-2p$ core levels at $250$ eV and $350$ eV carried out on the same spot for both terminations. We observe intensity differences between the two photon energies for both terminations. Notably, intensity for the main peaks at $\sim163.1$ eV and $\sim161.8$ eV are both amplified while the shoulders at $\sim162.6$ eV and $\sim161.3$ eV in panel a ($\sim162$ eV and $\sim160.7$ eV for panel b) are both mildly diminished with increasing photon energy. The calculated Inelastic Mean-free path using the TPP-2M\cite{TPP2m} formula at the two photon energies are $4.66$ \AA \, and $5.94$ \AA \, respectively Based on the normalization, this difference can be attributed to the different depth sensitivity at both these photon energies. %With greater photon energy, the kinetic energy ( of the photoemitted electrons are larger allowing more electrons from deeper in the sample to escape the material. This would result in a pronounced amplification of the peaks that correspond to  bulk states as is seen with the main peaks, establishing the bulk origin.

\section{\large Supplementary Note 3: Additional Details for SESSA simulation}

\begin{figure}
    \centering
    \includegraphics[width=0.5\linewidth]{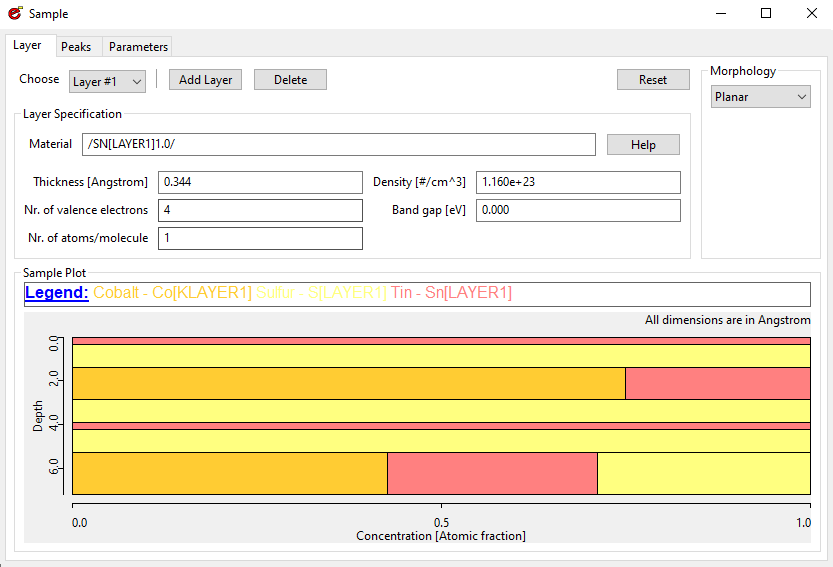}
    \caption{SESSA sample model. Atomic Densities were based on actual atomic densities of the layers in the crystal structure}
    \label{SESSA_sample}
\end{figure}

Figures 2 and 8 in the main paper use SESSA to simulate XPS spectra. A layered model for the crystal was used for the simulation, as shown in Fig. \ref{SESSA_sample}. Actual atomic densities for the material, from the crystal structure, were used. Atomic species were distinguished into surface and bulk, with only the surface atoms being given a chemical shift based on DFT calculations. Lorentzian line shapes with $0.22$ eV widths were used for all peaks, as discussed in methods.

\begin{figure}
    \centering
    \includegraphics[width=0.5\linewidth]{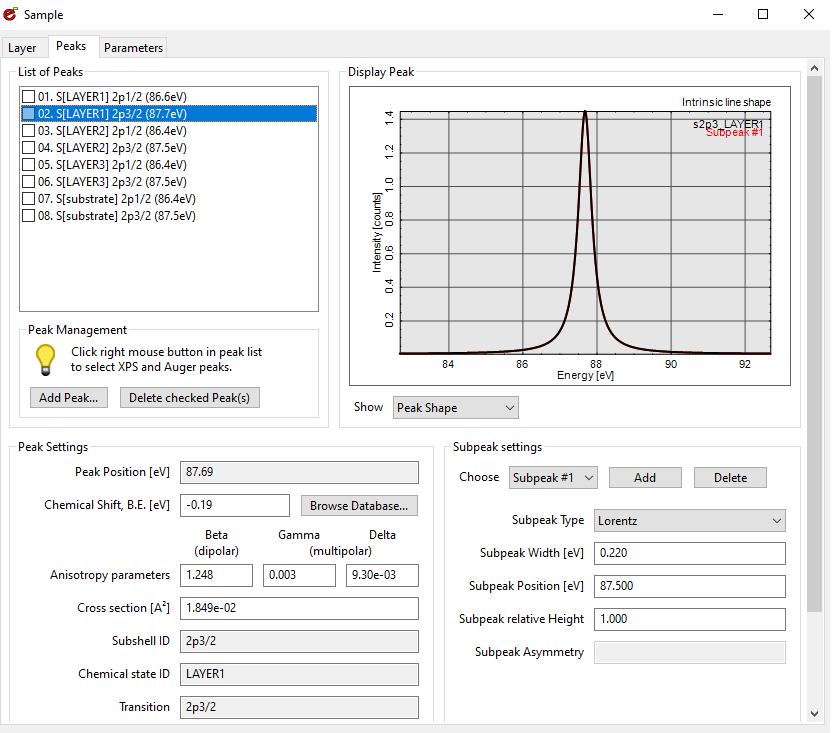}
    \caption{SESSA peaks panel containing details of peak offsets (from DFT) and line widths used}
    \label{SESSA_peaks}
\end{figure}

\section{\large Supplementary Note 4}
\begin{figure}
    \centering
    \includegraphics[width=0.5\linewidth]{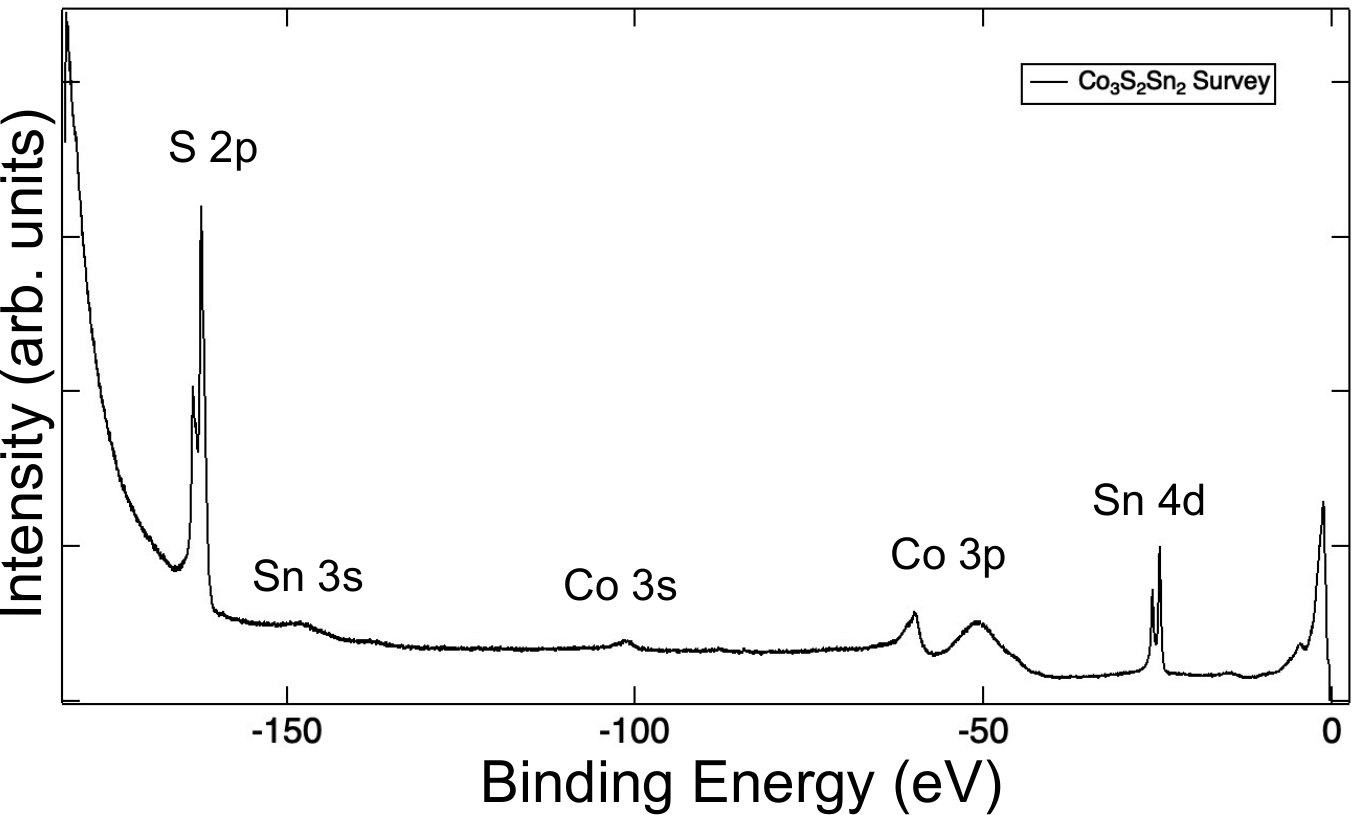}
    \caption{Survey of core levels in Co$_3$Sn$_2$S$_2$ at photon energy of $200$ eV}
    \label{corelevels200}
\end{figure}

\begin{figure}
    \centering
    \includegraphics[width=0.5\linewidth]{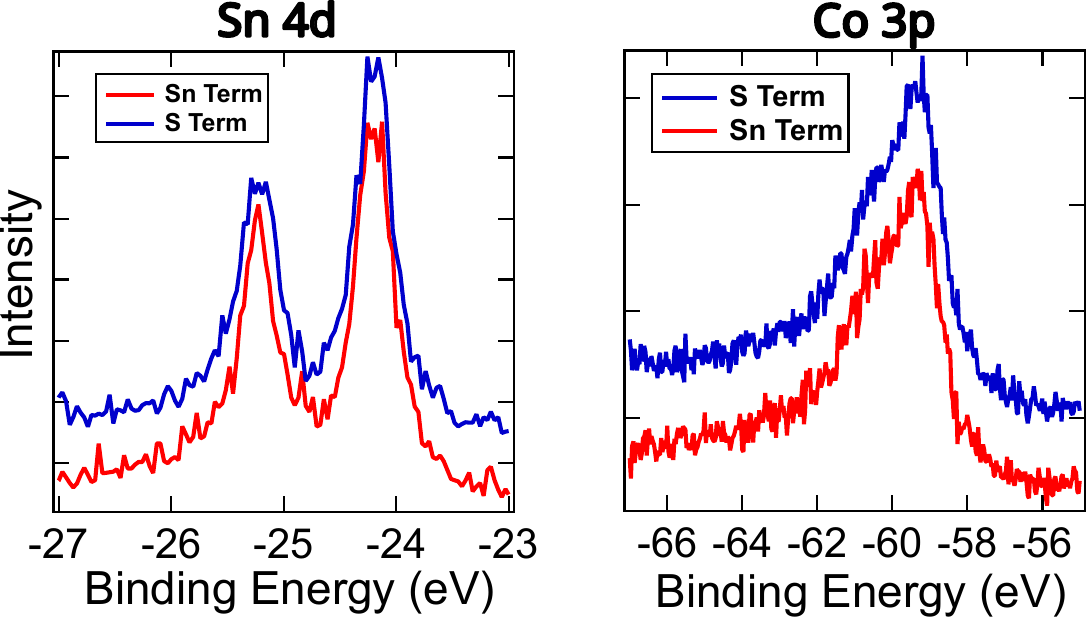}
    \caption{Sn$-4d$ and Co$-3p$ core levels taken at $250$ eV}
    \label{corelevels250}
\end{figure}

Fig. \ref{corelevels200} and \ref{corelevels250} show other relevant core levels that are not discussed in the main text where we focus on the information-rich S$-2p$ core level. Fig. \ref{corelevels200} shows a broad survey taken at $200$ eV, which shows core levels corresponding to all elements in the compound. Further investigation at $250$ eV of the Sn$-4d$ and Co$-3p$ levels (Fig. \ref{corelevels250}) indicates minimal differences at the two terminations as discussed in main text.

\section{\large Supplementary Note 5}
\begin{figure}
    \centering
    \includegraphics[width=0.3\linewidth]{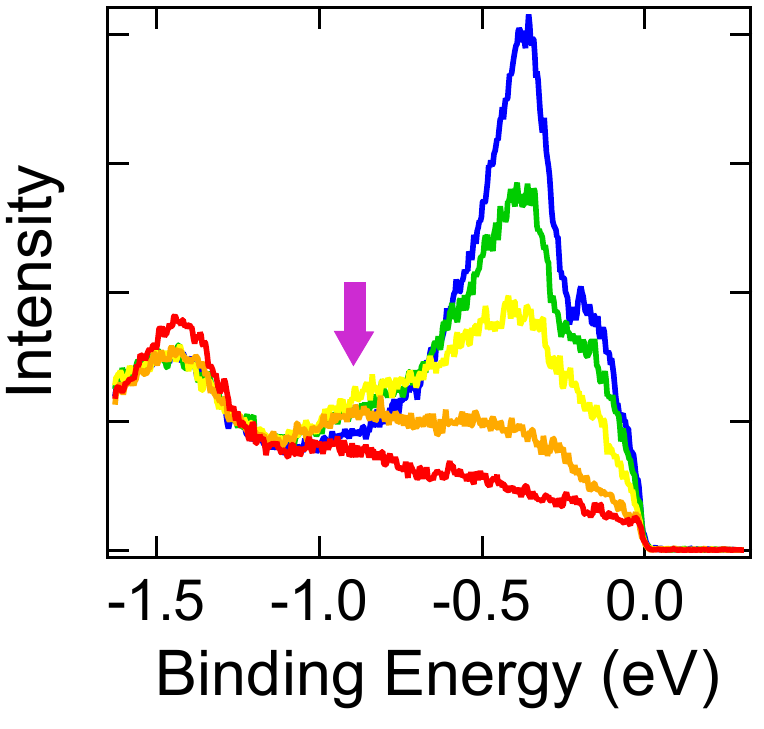}
    \caption{Raw EDC at $\Gamma$ for spectra shown in Fig 6 of main text. Colors are consistent with main text figure. Arrow indicates additional spectral weight not found in ARPES spectrum of pure terminations}
    \label{gamma_edc}
\end{figure}

The main text references disorder-related effects on the presence of features exclusive to the intermediate region. We illustrate this in Fig. \ref{gamma_edc} with the EDCs at the $\Gamma$ point for the same spatial points in Fig 6 of the main text. The raw EDCs of the intermediate region (Orange, Yellow and Green) very clearly demonstrate a pronounced increased intensity at approximately $0.85$ eV binding energy. This highlighted feature manifests in the raw spectrum as a 'vertical' feature at normal emission, as discussed in main text.  The prominent peak at around  $0.4$ eV binding energy in blue, green, and yellow EDCs is the distinct surface feature of the S termination, which is seen to become broader and weaker intensity in the intermediate region.

\section{\large Supplementary Note 6}

 \begin{figure*}
    \centering
    \includegraphics[width=0.8\textwidth]{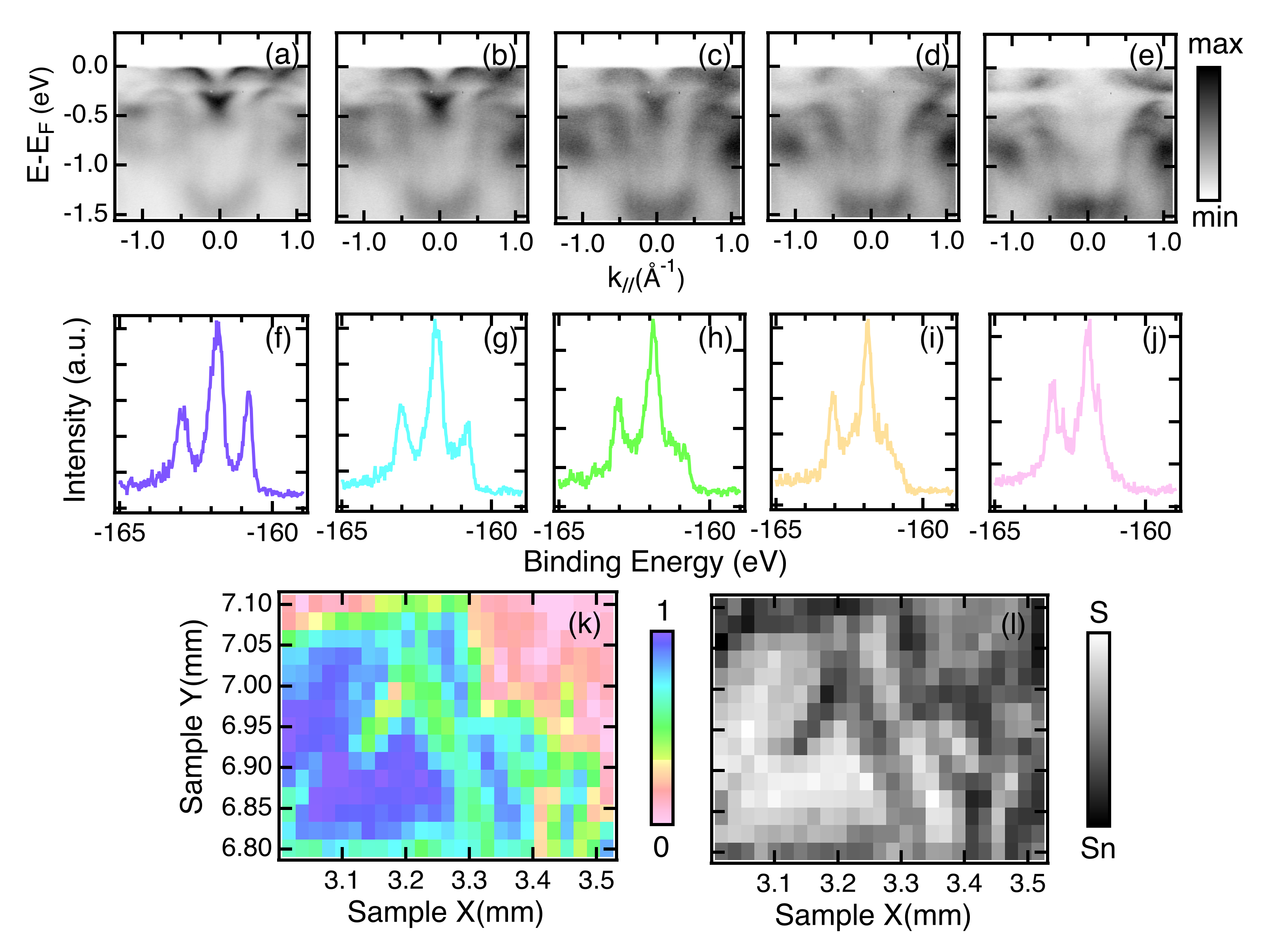}
	\caption{
     Global fitting of EDCs at M point to linear combination of pure termination EDCs. Top panel (a)-(e), ARPES spectra from sites with fitting parameter ($a$ in equation \ref{spatial_fit_eqn}) 1, 0.75, 0.5, 0.25, 0 respectively. Each is an average of 5 spots with errors in the fitting parameter within $\pm$0.02. Middle panel (f)-(j), Averaged S$-2p$ core level XPS spectra from the same spots as top panel. Color of each also corresponds to the map in the bottom panel. (k)Spatial variation of the fitting parameter. (l) Spatial
map of S$-2p$ S-termination feature, same as in main text Fig 3(a).
	}
	\label{fitmap}
\end{figure*}

Fig 6 in the main text shows the results of a series of EDC fits to a few selected points in the spatial map at the M' point in the phase space. The fitting was conducted under the assumption that the EDC at each site consists of a linear combination of pure S-termination EDC and pure Sn-termination EDC. To further illustrate this idea, we demonstrate the result of the same fitting on the entire spatial map in Fig. \ref{fitmap}. The pure terminations we selected using the UMAP analysis of Fig 3 from the main paper to find the 'purest' S and Sn terminations. Each EDC in the spatial data set is normalized to its area in the (-0.5,0) eV binding energy range and then fitted each to a linear sum as:

\begin{equation}
a\cdot \text{EDC}_{\text{S}}+(1-a)\cdot\text{EDC}_{\text{Sn}} 
\label{spatial_fit_eqn}
\end{equation}

Examples of these fits are shown in Fig. 6 of the main text.

The top (a-e) and middle panels (e-j) of Fig. \ref{fitmap} show the averaged ARPES and XPS spectra from points with different fitting parameters, which is essentially grouping different points in space by how much S content they have. In these averaged spectra, we can see the sulfur surface state feature at $\Gamma$ diminishing as the fitting parameter goes to zero.  These ARPES spectra are comparable to spectra from the intermediate region in Fig. 6b of the main text, though it should be noted that they are produced by averaging several spectra at different positions, rather than being individual spectra. Similarly, averaged S$-2p$ core levels in Fig. \ref{fitmap} (f)-(j) show an evolving surface feature at low binding energy.

\begin{figure*}
    \centering
    \includegraphics[width=0.6\textwidth]{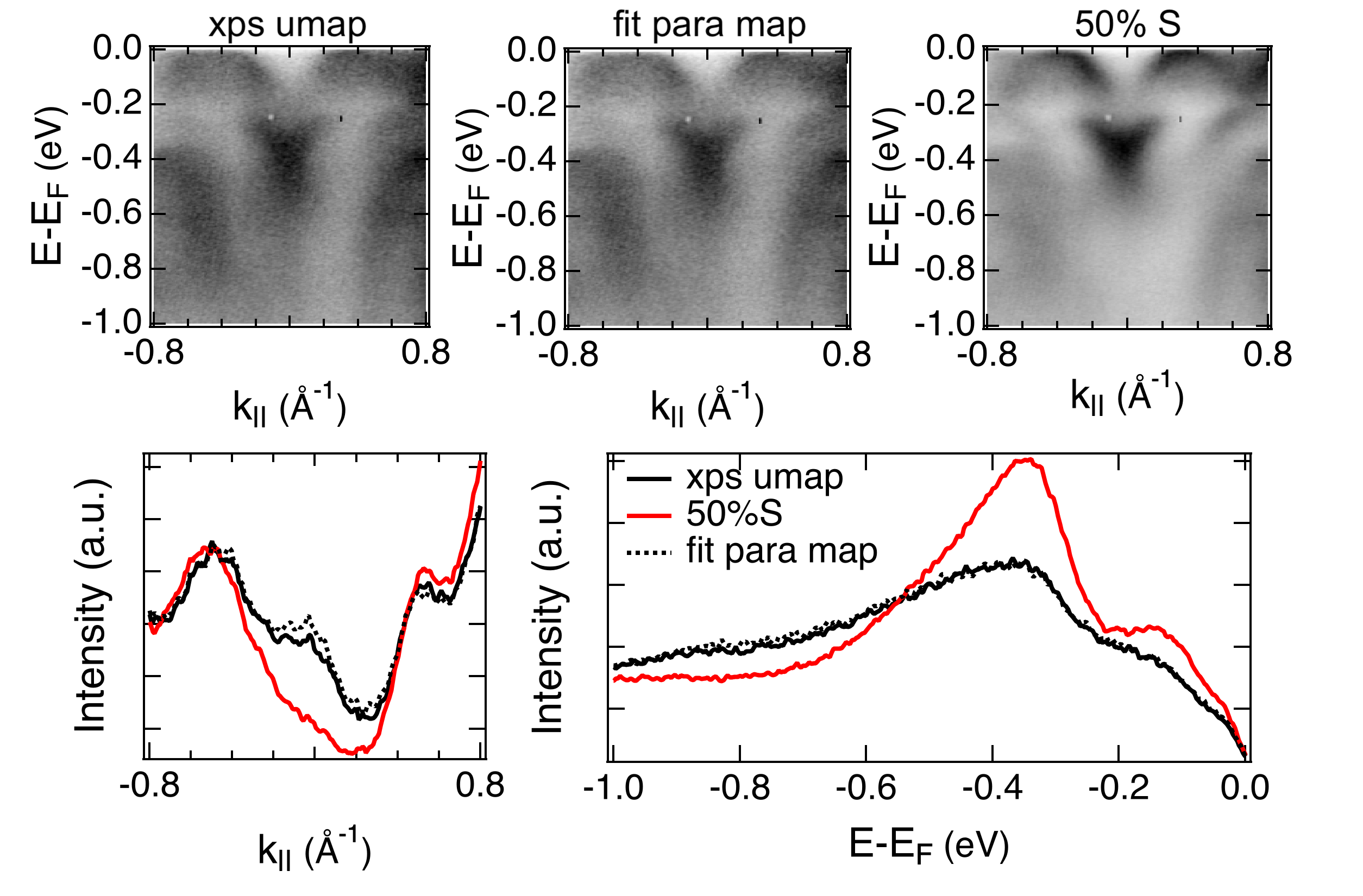}
    \caption{Comparison between intermediate region measured spectra and reconstructed linear combination fit, with two ways of selecting comparable measured spectra. Measured spectrum averaging five points chosen via: (a) XPS UMAP (main text Figure 5(b)) with nearly identical UMAP coordinates (same procedure as main text) (b) linear combination fit parameter map(Fig.\ref{fitmap}) with value $0.498\pm0.015$, (c) Reconstructed spectrum using a linear combination of 49.8\% S termination and 50.2\% Sn termination, (d) MDCs from each spectrum taken at $E_B$= -0.8 eV, (e) EDCs taken at $\Gamma$ point}.
    \label{comparison}
\end{figure*}

\section{\large Supplementary Note 7}

\begin{figure*}[h]
    \centering
    \includegraphics[width=0.8\textwidth]{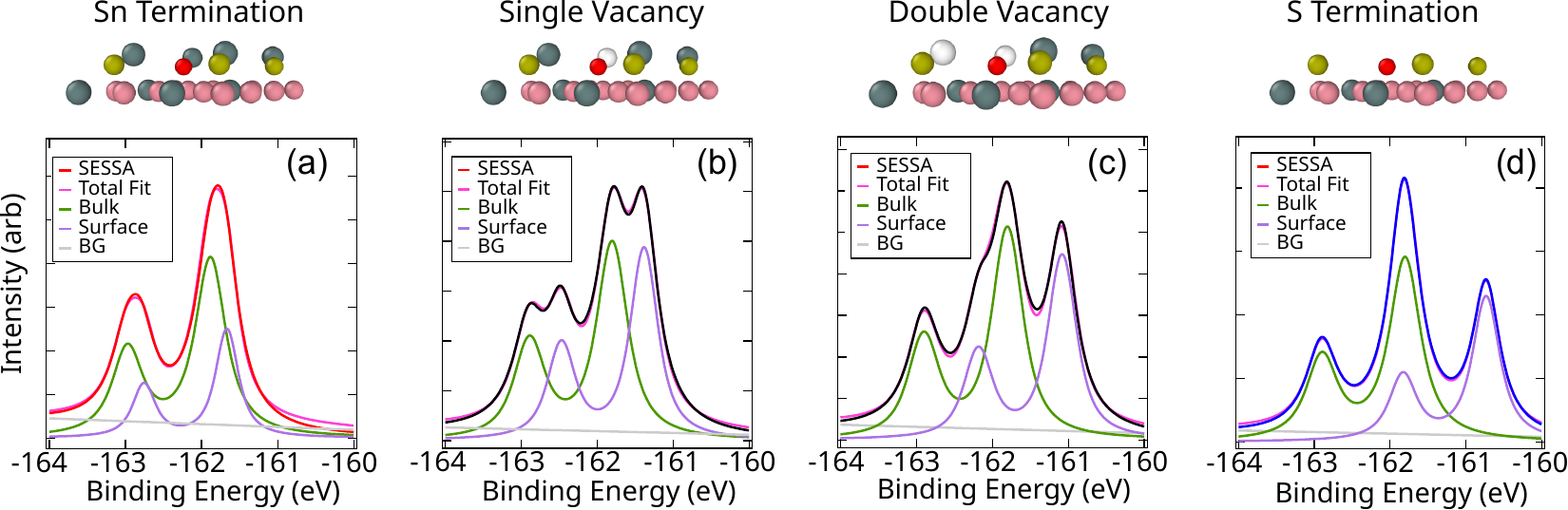}
    \caption{Simulations for different configurations of defects showing bulk S$-2p$ core level plus surface contribution from a single type of defect. a) Pure Sn termination b) Single vacancy c) Double vacancy d) Pure S termination.  Chemical shift from DFT as discussed in methods.}
    \label{sim}
\end{figure*}

We also simulated core level spectra for the S$-2p$ orbitals for different configurations of the surface vacancy (Fig. \ref{sim}). DFT-calculated surface core level shifts were used for the SESSA simulation. We can clearly see the decreasing binding energy as we transition from the Sn termination to the S termination. It must be noted that pure vacancy terminations are hard to observe in photoemission spectra because of the macroscopic spot size. Thus, a superposition of each of these spectra is more likely observed.

\section{\large Supplementary Note 8}

\begin{figure*}[htb]
    \centering
    \includegraphics[width=0.8\textwidth]{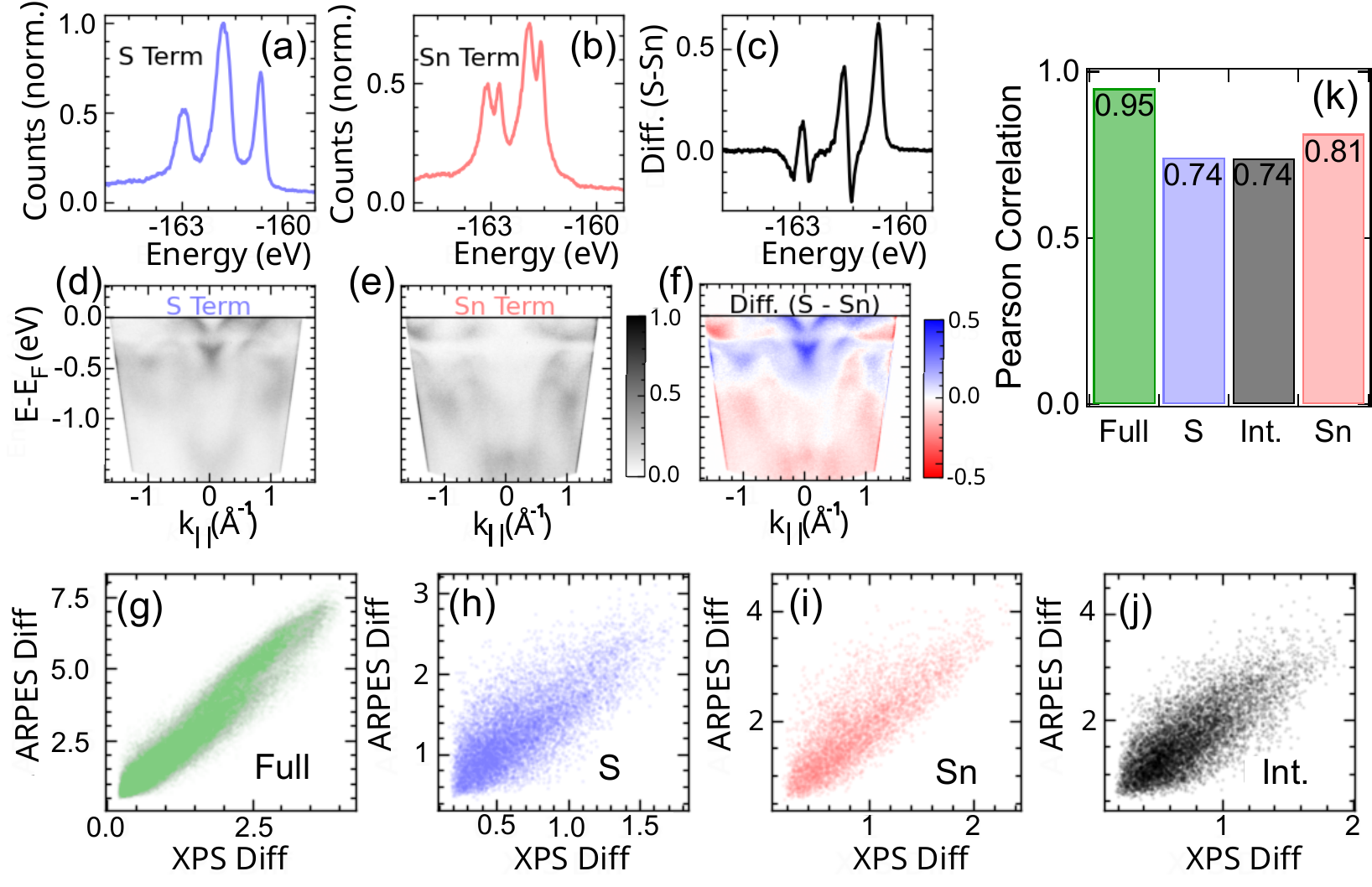}
    \caption{Correlation between local S$-2p$ core levels and ARPES spectra. Example core levels for S (a) and Sn (b) termination as well as their difference, normalized to background (c). ARPES spectra for S (d) and Sn (e) termination and their difference (f). Scatter plots of XPS and ARPES difference metrics for the full data set (g), as well as S (h), Sn (i), and mixed (j) cluster subsets. (k) Pearson correlation between ARPES and S$-2p$ core level difference spectra for each of the previous clusters and the entire data set. Numbers are obtained values}
    \label{cluster}
\end{figure*}

Correlations between core levels and ARPES are shown in the main text and discussed in more detail here in Fig. \ref{cluster}.  To quantify the correlation between XPS and ARPES spectra, the following procedure is employed. First, all spectra were normalized to the background to account for differing photoemission intensity. For the core levels, this was done by dividing the spectra by the difference between intensity at the high and low binding energy tails outside the S$-2p$ peak regions. For the ARPES spectra, integrated intensity above $E_F$  was used to divide each spectra. Second, spectral differences for every pair of spatial points was calculated using the normalized spectra. An example of this difference for the core levels is shown in Fig. \ref{cluster}(a-c), wherein the difference (c) between spectra on two pure terminations S (a) and Sn (b) is demonstrated. The difference is also computed for the respective ARPES spectra in panels (d)-(f). Finally, each pairwise difference is reduced to a single number with a normalized absolute difference. These pairwise differences are summarized in Fig. \ref{cluster}(g)-(j) , which show scatter plots of spectral differences for every pair of spatial points, where the x-axis is the difference in the XPS channel and the y-axis is the difference in the ARPES channel.  The correlation within these scatter plots is then quantified using the Pearson correlation coefficient, r (Fig. \ref{cluster}(k) ). A coefficient close to one indicates that data points exist on a tight diagonal where a small difference between two XPS spectra always implies a small difference between ARPES spectra and vice versa. The calculated Pearson coefficients for the full dataset, S, intermediate, and Sn termination are 0.95, 0.74, 0.74 and 0.81 respectively.  This analysis suggests strong correlation between the local S-$2p$ core levels and local ARPES spectra, especially when considering the entire data set.  In the manuscript, this correlation is used to identify 'similar' ARPES spectra using simpler XPS spectra.  However, it should be noted that XPS spectra are not always determinative of ARPES spectra.  For example, in Fig. \ref{corelevels250}, we show that Sn-4d and Co-3p core levels on this same material show minimal dependence on the termination, despite Sn being one of the terminations observed.  Moreover, ARPES spectra can be affected by surface imperfections that do not affect core levels.  This includes sample curvature, cleave-induced surface mosaicity, surface roughness, and stacking differences in 2D materials.

\bibliographystyle{unsrtnat}
\bibliography{references}